\documentclass[particles,article,accept,pdftex,moreauthors]{Definitions/mdpi} 

\firstpage{1} 
\pubvolume{1}
\issuenum{1}
\articlenumber{0}
\pubyear{2022}
\copyrightyear{2022}
\externaleditor{Academic Editor: {Armen Sedrakian} 
}
\datereceived{} 
\dateaccepted{} 
\datepublished{} 
\hreflink{https://doi.org/} 

\Title{Quick Guides for Use of the CompOSE Data Base}

\TitleCitation{{Quick Guides} 
 for Use of the CompOSE Data Base}

\Author{{Veronica Dexheimer} 
 $^{1}$\orcidA{}, Marco Mancini $^{2,3}$, Micaela Oertel $^{3}$\orcidB{}, {Constança Provid\^encia} 
 $^4$,  Laura Tolos $^{5,6,7}$ \linebreak and {Stefan Typel} 
 $^{8,9,}$*\orcidC{}}

\AuthorNames{Veronica Dexheimer, Marco Mancini, Micaela Oertel, Constanca Provid\^encia,
Laura Tolos and Stefan Typel}

\AuthorCitation{{Dexheimer, V.;} 
 Mancini, M.; Oertel, M.; Provid\^encia, C.; Tolos, L.; Typel, S.}

\address{%
$^{1}$ \quad Department of Physics, Kent State University, Kent, OH 44242, USA; vdexheim@kent.edu\\
$^{2}$ \quad {Laboratoire} 
 IDP---C.N.R.S. UMR 7013---{Institut Denis-Poisson,} 
 Université d’Orléans-Université de Tours, {France;} 
 marco.mancini@obspm.fr\\
$^{3}$ \quad LUTH, {Observatoire de Paris,} 
 Universit\'e PSL, CNRS, Universit\'e Paris Cité, 92190 Meudon, France; micaela.oertel@obspm.fr\\
$^{4}$ \quad {CFisUC,} 
 Department of Physics, University of Coimbra, 3004-516 Coimbra, Portugal; cp@uc.pt\\
$^{5}$ \quad Institute of Space Sciences (ICE, CSIC), Campus UAB,  Carrer de Can Magrans, 08193 Barcelona, Spain; tolos@ice.csic.es\\
$^{6}$ \quad Institut d'Estudis Espacials de Catalunya (IEEC), 08034 Barcelona, Spain \\
$^{7}$ \quad Frankfurt Institute for Advanced Studies, Ruth-Moufang-Str. 1, 60438 Frankfurt am Main, Germany\\
$^{8}$ \quad {Technische Universit\"{a}t Darmstadt, Fachbereich} 
 Physik, Institut f\"{u}r Kernphysik, Schlossgartenstra\ss{}e 9, \linebreak 64289 Darmstadt, Germany; stypel@ikp.tu-darmstadt.de\\
$^{9}$ \quad GSI Helmholzzentrum f\"{u}r Schwerionenforschung, Theorie, Planckstra\ss{}e 1, 64291 Darmstadt, Germany}
\corres{Correspondence: {develop.compose@obspm.fr} 
}

\abstract{We present a combination of two quick guides aimed at summarizing relevant information about the \textsc{CompOSE} nuclear equation of state {repository}. The first is aimed at nuclear physicists and describes how to provide standard equation of state tables. The second quick guide is meant for users and describes the basic procedures to obtain customized tables with equation of state data.  Several examples are included to help providers and users to understand and benefit from the \textsc{CompOSE} database.}

\keyword{\textsc{CompOSE} repository; users quick guide; providers quick guide} 

\begin{document}


\section{Introduction: The CompOSE Data Base}
The main purpose of the {online service} 
\textsc{CompOSE} (\url{https://compose.obspm.fr})
is to provide information and data for different equations of states (EoSs)
ready for further use in astrophysical applications, nuclear physics,
and beyond, see the review \cite{Oertel:2016bki} for a general
introduction. To that end, \textsc{CompOSE} contains a repository of
EoS data in tabulated form following a common format with information
on a large number of thermodynamic properties and, if available, on
the chemical composition of dense matter and on microphysical
properties of the constituents. \textsc{CompOSE} not only allows
direct downloading of data together with a bibliography on data-related
publications, but also provides several tools to handle and customize data,
for instance for extracting selected quantities, interpolation of tabulated data, or calculation of
additional related quantities. A
full documentation with detailed instructions can be found in the full
manual~\cite{manual:1,manual:2}. Here, we present quick guides for potential providers of EoS
data and for users of
the \textsc{CompOSE} service.  One word
of caution is in order: \textsc{CompOSE} is designed to evolve and to
be extended over time; thus, the presentation below shows the current
status (July 2022) and will not be adapted in the future upon the adding of
new features to the service.

If you make use of the tables provided by \textsc{CompOSE}, please
cite the publications describing the respective EoS models (available
on the \textsc{CompOSE} web page for each EoS data table),
together with the original \textsc{CompOSE} publications
\cite{Oertel:2016bki,manual:1,manual:2} and the \textsc{CompOSE}
{website} 
 \url{https://compose.obspm.fr}.

\section{Instructions for Providers of EoS Data}
The success of \textsc{CompOSE} depends on the support of nuclear physicists
providing their data. A collection of EoS models is already
incorporated in the \textsc{CompOSE} database, however, a larger set of
EoS from different models is highly desirable.  Please contact the
\textsc{CompOSE} core {team} 
 (\url{develop.compose@obspm.fr}) if you wish to contribute. 

\subsection{Preparation of Tables}

In the \textsc{CompOSE} database, the EoS is assumed to describe dense
matter in thermodynamic equilibrium, i.e., thermal and mechanical
equilibrium. In addition, it is assumed that all the constituents are
in chemical equilibrium with respect to reactions mediated by the
strong and electromagnetic interactions. Whether equilibrium with
respect to weak $\beta$-type reactions is assumed depends on the
particular table. Tables designated for application to core-collapse
supernovae or binary neutron star mergers do not, whereas the tables
constructed to describe cold neutron star matter do, reducing in this
case the number of independent particle number densities,
i.e., thermodynamic parameters. For EoS models with strangeness-bearing
particles, it is assumed that strangeness-changing weak interactions
have enough time to equilibrate, in which case there is no constraint
related to the strangeness density and the strangeness chemical
potential vanishes. Except for the tables of pure hadronic and/or
quark matter (without leptons), charge neutrality is assumed to
hold. Neutrinos are never included in the present tables, since they
are usually treated independently from the EoS in astrophysical
simulations, not assuming thermodynamic equilibrium. Photons can be
included in finite-temperature tables. In the case of EoS for
pure-neutron matter, however, it is supposed that photons are not
included.

Quantities in \textsc{CompOSE} are given in natural units\index{units}
$\hbar=c=k_{B}=1$\index{$\hbar$}\index{$c$}\index{$k_{B}$} (for
details on unit conversion see the {NIST} 
 (\url{https://physics.nist.gov/cuu/Constants/index.html}) or {CODATA} 
 (\url{https://www.codata.org}) websites). Particle number
densities of all particles $i$ are given by $n_{i}\index{$n_{i}$} =
{N_{i}}/{V} \ [\mbox{fm}^{-3}]$, where $N_{i}$ is the particle number
inside the volume $V$. For particles with half-integer spin at finite
temperature, $n_{i}$ represents the net particle density, i.e.,\ the
difference between the number density of particles and antiparticles
(see full CompOSE
manual~\cite{manual:1,manual:2} for the possibility of entering particles and
antiparticles separately). For particles with integer spin, particle
and antiparticle number densities are given separately. The baryon
number density $n_{B}$ is given by $n_{B} = {N_{B}}/{V} = \sum_{i}
B_{i} n_{i} \ [\mbox{fm}^{-3}]$, where $N_{B}$ is the total baryonic
number and $B_{i}$ the baryon number of a given particle, e.g., the
baryon number for a quark is $1/3$.

The hadronic (and quark) charge density is given by
$n_{q}\index{$n_{q}$} = Q/V= \sum_{i}^{\prime} Q_{i} n_{i}
\ [\mbox{fm}^{-3}]$, where $Q$ is the total electric charge and
$Q_{i}$ is the electric charge of a given particle. The prime
indicates that the summation excludes leptons. For a nucleus
${}^{A_{i}}Z_{i}$, the baryon number and electric charge are the mass
number $A_{i}$ and atomic number $Z_{i}$, respectively. Particle
fractions are defined as $Y_{i}= n_{i}/n_{B}$ [\mbox{dimensionless}].

The hadronic (and quark) charge fraction, from hereon simply charge
fraction, is defined as $Y_{q}={n_{q}}/{n_{B}}$
$[\mbox{dimensionless}]$. In models with electrons and muons, charge
neutrality requires $ Y_{q} = Y_{e}+Y_{\mu} = Y_{l}\index{$Y_{l}$}$\,,
where $Y_{l}$ is the lepton fraction (not considering
neutrinos). Because of the imposed physical conditions, the state of
the system is uniquely characterized by the three quantities
temperature $T$ [MeV], baryon number density
$n_{B}$ [fm$^{-3}$], and charge fraction
$Y_{q}$. The latter variable is used because it is also
defined in pure hadronic EoS models that do not include charged
leptons.

Table \ref{tab1} introduces an indexing scheme for the most
relevant particles that identifies them uniquely (more can be added
upon request---see the full manual). For a nucleus ${}^{A}Z$, the index is
$1000\cdot{A}+Z$ and, for the photon $\gamma$, the index is 600. Note
that it should be stated in the accompanying information if a given
EoS includes photon contributions. This should appear in a pdf file
(the so-called `data sheet') that includes a short characterisation of
the EoS~model, relevant references, the meaning of an index
$I_{\textrm{phase}}$ for the phases that appear in the tables,
considered particle species, parameter ranges, additional quantities
provided,~etc.
\begin{table}[H]
\caption{{Particle} 
 indices. \label{tab1}}
\newcolumntype{C}{>{\centering\arraybackslash}X}
\begin{tabularx}{\textwidth}{CCCCCCCCCCCC}
\toprule
$e^{-}$&$\mu^{-}$&$n$&$p$&$\Delta^{-}$&$\Delta^{0}$&$\Delta^{+}$&$\Delta^{++}$&$\Lambda$&$\Sigma^{-}$&$\Sigma^{0}$&$\Sigma^{+}$\\
\midrule
0&1&10&11&20&21&22&23&100&110&111&112\\
\midrule

$\rho^{-}$&$\rho^{0}$&$\rho^{+}$&$\delta^{-}$&$\delta^{0}$&$\delta^{+}$&$\pi^{-}$&$\pi^{0}$&$\pi^{+}$&$\phi$&$\sigma_{s}$&$K^{-}$\\
\midrule
300&301&302&310&311&312&320&321&322&400&410&420\\
\midrule

$\Xi^{-}$&$\Xi^{0}$&$\omega$&$\sigma$&$\eta$&$\eta^{\prime}$&$K^{0}$&$\bar{K}^{0}$&$K^{+}$&$u$&$d$&$s$\\
\midrule
120&121&200&210&220&230&421&422&423&500&501&502\\
\bottomrule
\end{tabularx}
\end{table}

\subsection{EoS Tables and Data Format}

There are at least four files with numerical data required for an EoS
to be included in the \textsc{CompOSE} repository: three files that
specify the discretization scheme of the independent variables
temperature, baryon number density, and charge {fraction} 
 ({\tt eos.t},
{\tt eos.nb}, and {\tt eos.yq}), and a file ({\tt eos.thermo}) with the
table of thermodynamic quantities. Additional data on the chemical
composition and on microscopic quantities can be collected in two
additional files ({\tt eos.compo} and {\tt eos.micro}).

The variables $T$, $n_{B}$, and $Y_{q}$ are given on a
provider-defined grid in which each point is identified by three
indices. The values corresponding to these indices are given in
separate files (and should ideally have at least eight significant
digits):
\begin{itemize}
\setlength\itemsep{0pt}
\item \textbf{{temperature} 
 $T$} [MeV] in file {\tt eos.t}, recommended to increase logarithmically (at least at large temperatures),
\item \textls[-30]{\textbf{{baryon number density} $n_{B}$} [fm$^{-3}$] in file {\tt eos.nb}, recommended to increase~logarithmically,}
\item \textbf{{charge fraction} $Y_{q}$} [dimensionless] in file {\tt eos.yq}, recommended to increase linearly in~$Y_q$.
\end{itemize}

In Section~\ref{sec:ranges} we list recommended ranges of these
parameters for the different types of tables. In these files, the
first two lines should be the minimum and maximum indices
(representing the respective minimum and maximum values that are
given). The following lines give the numerical values of the variables
for the indices in ascending order. Each line of the data files {\tt
  eos.thermo}, {\tt eos.compo}, and {\tt eos.micro}, except the first
line in {\tt eos.thermo}, starts with the indices corresponding to the
three variables in the order $T$, $n_{b}$, $Y_{q}$. In the particular
case of $\beta$ equilibrium and zero temperature, the files {\tt
  eos.t} and {\tt eos.yq} are {simply} 
\vspace{.2cm}\par \hspace{3cm}0 \hspace{5.35cm} 1 
\par \hspace{3cm}0 \hspace{5.35cm} 1
\par \hspace{3cm}0.00000000E+00  \hspace{3cm} 0.00000000E+00

\subsubsection{Thermodynamic Properties}

These are listed in the file {\tt eos.thermo}. It contains three entries in the
first line, namely, the masses of the neutron and proton in MeV,
and an integer that indicates if the EoS contains leptons (1) or not
(0). The remaining lines contain the following entries
\begin{equation}
 i_{T} \:\:\: i_{n_{B}} \:\:\: i_{Y_{q}} \:\:\: \frac{p}{n_{B}} \:\:\: s \:\:\: \frac{\mu_{b}}{m_{n}}-1 \:\:\: \frac{\mu_{q}}{m_{n}} \:\:\: \frac{\mu_{l}}{m_{n}} \:\:\: \frac{f}{n_{B}m_{n}}-1 \:\:\:
 \frac{e}{n_{B}m_{n}}-1 \:\:\:
 N_{\textrm{add}} \:\:\: 
 \underbrace{q_{1} \:\:\: 
   q_{2} \:\:\: \dots
 \:\:\: }_{N_{\textrm{add}} \:\: {\textrm{quantities}}}\nonumber
\end{equation} 
corresponding to the indices for temperature, baryon number density, and charge fraction, followed by seven mandatory thermodynamic quantities. These are
\begin{itemize}
    \item pressure divided by baryon number density $p/n_{B}$ (MeV),
    \item entropy per baryon (or entropy density per baryon number density) $s$,
    \item scaled and shifted baryon chemical potential $\mu_B/m_n - 1$,
    \item scaled charge chemical potential $\mu_q/m_n$,
    \item scaled effective lepton chemical potential (set to zero in models without leptons) $\mu_l/m_n$,
    \item scaled and shifted free energy per baryon $f/(n_B m_n)-1$,
    \item scaled and shifted energy per baryon $e/(n_B m_n) -1$.
\end{itemize}

An integer follows, indicating the absence (0) or presence ($N_{add}$)
of $N_{add}$ additional optional thermodynamic quantities to be specified
in the data sheet. If the table contains repeated rows with identical
indices $i_{T}$, $i_{n_{B}}$, $i_{Y_{q}}$\,, only the last row is
used.

From thermodynamic identities, the free energy density is
$f(T,n_{B},Y_{q}) = -p + \sum_{i} \mu_{i} n_{i}$, with
particle chemical potentials
$\mu_{i}=B_{i}\mu_{B}+Q_{i}\mu_{q}+L_{i}\mu_{l}$ and lepton numbers
$L_{i}$ that do not distinguish between electrons and muons. In pure
hadronic (and quark) models without leptons, it can be written as
$f(T,n_{B},Y_{q}) = -p + \left( \mu_{B} + Y_{q} \mu_{q}
\right) n_{B}$. In the case that charged leptons are present, charge
neutrality requires $Y_{l} = Y_{q}$. The condition of (neutrinoless)
$\beta$ equilibrium corresponds to $\mu_{l}=0$ (assuming identical
lepton chemical potentials of electrons and muons $\mu_{e^{-}} =
\mu_{\mu^{-}} = \mu_{l}-\mu_{q}$) and $f(T,n_{B}) = -p + \mu_{B} n_{B}$.

\subsubsection{Composition of Matter (Optional)}

This is listed in the file {\tt eos.compo} containing in all lines the entries
\begin{eqnarray}
  \lefteqn{i_{T} \:\:\: i_{n_{B}} \:\:\: i_{Y_{q}} \:\:\: 
 I_{\textrm{phase}} \:\:\: N_{\textrm{pairs}} \:\:\: 
 \underbrace{I_{1} \:\:\: Y_{I_{1}} \:\:\: \dots \:\:\: 
}_{N_{\textrm{pairs}} {\textrm{pairs}}}} \hspace{6.4cm}
 N_{\textrm{quad}} \:\:\:
 \underbrace{I_{1} \:\:\:  A^{\textrm{av}}_{I_{1}}\:\:\:
   Z^{\textrm{av}}_{I_{1}} \:\:\: Y_{I_{1}} \:\:\: \dots \:\:\: 
   }_{N_{\textrm{quad}} \:\:\: {\textrm{quadruples}}} \nonumber
\end{eqnarray}
corresponding again to the indices for temperature, baryon number
density, and charge fraction, followed by an index encoding the type
of phase (chosen by provider and identified in the data sheet), number
of particles (pairs) for which the composition is given, particle
pairs in no particular order (particle indices from Table \ref{tab1} followed
by the respective particle charge fractions), number of particle
quadruples, particle quadruples (particle indices from Table \ref{tab1}
followed by the respective average mass and charge numbers of the
representative nucleus and respective combined charge fraction). In
each quadruple, the index $I_{i}$ specifies a group of nuclei
$\mathcal{M}_{I_{i}}$ with average mass number $
A^{\textrm{av}}_{I_{i}} = {\sum_{j\in \mathcal{M}_{I_{i}}}
  (A_{j}Y_{j}})/{\sum_{j\in \mathcal{M}_{I_{i}}} Y_{j}}$, average
charge number $Z^{\textrm{av}}_{I_{i}} = {\sum_{j\in
    \mathcal{M}_{I_{i}}} (Z_{j} Y_{j}})/{\sum_{j\in
    \mathcal{M}_{I_{i}}} Y_{j}}$, and combined charge fraction
$Y_{I_{i}} = \sum_{j\in \mathcal{M}_{I_{i}}} Y_{j}$. In the case that
there are no quadruples to report, $N_{\textrm{quad}}=0$. The
correlation between $I_{i}$ and $\mathcal{M}_{I_{i}}$ should
appear in the data sheet.
 
\subsubsection{Stellar Information (Optional)}

This is listed in the file {\tt eos.mr} containing in the first line the provided quantities (with units) preceeded by a ``\#''  sign and in the remaining lines
\begin{eqnarray}
 & &
 R \:\:\: M \:\:\:\Lambda \:\:\: n_c
\end{eqnarray}
corresponding to the radius (in km), mass (in solar masses), dimensionless tidal deformability, and central density (in fm$^{-3}$) of a family of cold $\beta$-equilibrated spherical neutron stars obtained from the provided equation of state. The third and fourth columns are optional and more relevant quantities can be added in additional columns.

\subsubsection{Microscopic Information (Optional)}

This is listed in the file {\tt eos.micro} containing in all lines the entries
\begin{eqnarray}
 & &
 i_{T} \:\:\: i_{n_{B}} \:\:\: i_{Y_{q}} \:\:\:
 N_{\textrm{qty}} \:\:\: 
 \underbrace{K_{1} \:\:\: q_{K_{1}} \:\: K_{2} \:\:
   q_{K_{2}} \:\:\: \dots \:\:\: }_{N_{\textrm{qty}} \: {\textrm{pairs}}} 
\end{eqnarray}
corresponding again to the indices for temperature, baryon number
density, and charge fraction, followed by the number of stored
quantities (pairs), the composite correlation indices that identify
uniquely the particle or correlation with the physical quantity $K_{i}
=~1000 \: I_{i} + J_{i}$ ({$I_{i}$} found in Tables \ref{tab1} and \ref{tab2}), and the
microscopic quantity index ($J_i$ found in Table \ref{tab3}). The microscopic
quantities available so far are the Landau mass $m^{L}_{i}$, the
effective Dirac mass $m^{D}_{i}$, the single-particle potential
$U_{i}$, the vector self-energy $V_{i}$, the scalar self-energy
$S_{i}$, and the size of superconductivity or superfluidity pairing
gaps $\Delta_{i}$ (see full manual for more details).

\begin{table}[H]
\caption{{Two-body} 
 correlation indices (with most relevant channel). \label{tab2}}
\newcolumntype{C}{>{\centering\arraybackslash}X}
\begin{tabularx}{\textwidth}{m{2cm}<{\raggedleft}m{3cm}<{\raggedleft}m{3cm}<{\raggedleft}m{3cm}<{\raggedleft}}
\toprule
\boldmath{$nn$ (${}^{1}S_{0}$)}& \boldmath{$np$ (${}^{1}S_{0}$)}&\boldmath{$pp$ (${}^{1}S_{0}$)}&\boldmath{$np$ (${}^{3}S_{1}$)}\\
\midrule
700&701&702&703\\
\bottomrule
\end{tabularx}
\end{table}
\vspace{-6pt} 

\begin{table}[H]
\caption{{Indices} 
 for microscopic quantities.}\label{tab3}
\newcolumntype{C}{>{\centering\arraybackslash}X}
\begin{tabularx}{\textwidth}{m{2.5cm}<{\raggedleft}m{2cm}<{\raggedleft}m{1.5cm}<{\raggedleft}m{1.5cm}<{\raggedleft}m{1.5cm}<{\raggedleft}m{1.5cm}<{\raggedleft}}
\toprule
\boldmath{$m^{L}_{I_{i}}/m_{I_{i}}$} & \boldmath{$m^{D}_{I_{i}}/m_{I_{i}}$} & \boldmath{$U_{I_{i}}$} & \boldmath{$V_{I_{i}}$} & \boldmath{$S_{I_{i}}$} & \boldmath{$\Delta_{I_{i}}$} \\ 
 \mbox{} \textbf{[Dimensionless] }& \textbf{[Dimensionless]} & \textbf{[MeV]} & \textbf{[MeV]} & \textbf{[MeV]} & \textbf{[MeV]}\\
\midrule
40 & 41 & 50 & 51 & 52 & 60\\
\bottomrule
\end{tabularx}
\end{table}

\subsubsection{Dimensionality of Tables}
\label{sec:ranges}
The recommended dimensions of the EoS data grids are:
\begin{itemize}
\setlength\itemsep{2pt}
    \item \textls[-35]{3D: {\bf {general-purpose EoS table}} with $N_{T}^{\textrm{max}} \times N_{n_{B}}^{\textrm{max}}
\times N_{Y_{q}}^{\textrm{max}} = 81  \times 301 \times 60=$ 1,462,860~data points, not including points with $T=0$~MeV when increasing logarithmically or $Y_{q}=0$. The temperature should start at $0.1$ MeV and the baryon number density at $10^{-12}$ fm${}^{-3}$ or~below,}
\item 2D: {\bf{{zero-temperature EoS table}}} with $N_{n_{B}}^{\textrm{max}} \times (N_{Y_{q}}^{\textrm{max}}+1) = 301 \times 61=$ 18,361 data points. The baryon number density should start at $10^{-12}$ fm${}^{-3}$ or below,
\item 2D: {\bf{{symmetric-matter EoS table}}} with 
$N_{T}^{\textrm{max}} \times N_{n_{B}}^{\textrm{max}} = 81 \times 301=$ 24,381 data points, not including points with $T=0$~MeV when increasing logarithmically all the way. The temperature should start at $0.1$ MeV and the baryon number density at $10^{-12}$ fm${}^{-3}$ or~below,
\item 2D: {\bf{{neutron-matter EoS table}}} with 
$N_{T}^{\textrm{max}} \times N_{n_{B}}^{\textrm{max}} = 81  \times 301$ = 24,381 data points, not including points with $T=0$~MeV when increasing logarithmically all the way. The temperature should start at $0.1$ MeV and the baryon number density at $10^{-12}$ fm${}^{-3}$ or~below,
\item 2D: {\bf{{EoS table of $\beta$-equilibrated matter}}} with 
$N_{T}^{\textrm{max}} \times N_{n_{B}}^{\textrm{max}} =81 \times 301$ =24,381 data points, not including points with $T=0$~MeV when increasing logarithmically all the way. The charge fraction is determined by the condition of charge neutrality and weak chemical equilibrium. The temperature should start at $0.1$ MeV and the baryon number density at $10^{-12}$ fm${}^{-3}$ or below,
\item 1D: {\bf{{EoS table of cold-symmetric matter}}} with 
$N_{n_{B}}^{\textrm{max}} = 301$ data points with $Y_{q}=0.5$ and $T=0$~MeV. The baryon number density should start at $10^{-12}$ fm${}^{-3}$ or below,
\item 1D: {\bf{{EoS table of cold-neutron}} matter} with 
$N_{n_{B}}^{\textrm{max}} = 301$ data points with $Y_{q}=0.0$ and $T=0$. The baryon number density should start at $10^{-12}$ fm${}^{-3}$ or below,
\item \textls[-25]{1D: {\bf{{EoS table of cold $\beta$-equilibrated matter}}} with $N_{n_{B}}^{\textrm{max}} = 301$ data points with \mbox{$T$ = 0 MeV}. The baryon number density should start at $10^{-12}$ fm${}^{-3}$ or below. Charge fraction 
is determined by the conditions of charge neutrality and weak chemical equilibrium. This table can be used directly by the {library} 
 \textsc{LORENE} (\url{https://lorene.obspm.fr}) to generate neutron-star models, and among others extract the maximum~mass. }
\end{itemize}

\section{Instructions for Users}

There are two equivalent ways to handle and customize data provided by
the \textsc{CompOSE} website. Both are
based on the {\tt compose} software, which is free and anyone can download it. A very
convenient way to handle and customize data is thus to use it
directly.  Alternatively, there is a web interface to the software,
accesible via the ``Compute'' button for each EoS table, which is free but password
restricted.  This quick guide describes the basic procedures to obtain
a table with EoS data with the help of a number of examples by direct
use of the software. 
It is prepared
in particular for first-time users in order to get familiar with the
files, the program and the handling. See Appendix \ref{app:a} for a summary of the notation we use. 

In this quick guide only the preparation of data tables in ASCII
format is described using a LINUX based operation system.  For an output of
data in the HDF5 format, please refer to the full CompOSE
manual~\cite{manual:1,manual:2}.

\subsection{Preparation of Program and Files}
\label{sec:code}
Download the compose code from the software section on the
\textsc{CompOSE} web page, where a link to the \textsc{CompOSE} gitlab
can be found ({To clone} 
 the git repository, please use the
  https protocol). You will need the files {\tt compose.f90},
  {\tt composemodules.f90} and {\tt Makefile}. Copy them to the same
directory.
Generate the executable by {typing} 
\begin{quotation}\noindent
{\tt make compose} \includegraphics[scale=1]{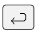}
\end{quotation}
The provided version of the {\tt Makefile} employs the {\tt gfortran} compiler, available on many LINUX systems. For another compiler, you have to modify the {\tt Makefile} accordingly. 

If during compilation an error related to HDF5 appears, check line 49 of the file {\tt Makefile}. Ensure that it  {reads}
  \begin{quotation}
    \noindent
    {\tt
      HDF5 = 0 
    }
  \end{quotation}

\subsubsection{EoS Data}

There are presently four families of EoS tables available in
\textsc{CompOSE}, i.e., General Purpose EoS, Neutron Matter EoS, Cold Neutron Star
EoS, and Cold Matter EoS.  You will see this option when clicking on
``EoS'' and then on ``Families'' in the left column of the \textsc{CompOSE} web
page.  The code should run with each of them. Alternatively, you can
find the EoS~model of your preferred author with the help of the
bibliography search on the web page. The files for download can be
found by clicking on 'Details' for each EoS~model. Download the files
for your preferred EoS table and move them to the same directory as the
code. Note that, by default, the data from all EoS~models have the same name. Thus, you should either remove all previous EoS data files before running a new model or create a different directory for each model.

\subsubsection{Input Parameters}

The code allows calculation of many different quantities, and surely not
every user needs all of them. In addition, some of them are only
available for selected EoS tables. Therefore, before running the code,
the grid in thermodynamic variables (temperature $T$, baryon number
density $n_B$, and charge fraction $Y_q$), as well as the desired
output quantities have to be specified. This is done with two input
parameter files, {\tt eos.parameters} for the grid and {\tt
  eos.quantities} for the output. Three examples, described below, are
available on the web page in the 'software' section.
  Note that if you do not remove the older versions of the input files
  {\tt eos.parameters} and {\tt eos.quantities}, you will be prompted
  to overwrite them upon creating new versions.

\subsubsection{Generating the Parameter Files}

Run the code by {typing}
\begin{quotation}
{\tt ./compose} \includegraphics[scale=1]{return.pdf}
\end{quotation}
in a terminal. You will then be prompted to select from three
options. Selecting task `1' will guide you to the generation of the
{\tt eos.quantities} file, specifying the needed output. Task~'2' will
guide you to the generation of the {\tt eos.parameters} file,
specifying the grid in thermodynamic variables. Just follow the
instructions. Keep in mind that the code is interpolating the EoS
data, it is thus not possible to calculate outside the ranges given by
your EoS data tables. You can check the ranges for each table on the
web page, or in the files {\tt eos.t, eos.nb}, and {\tt eos.yq}.

\subsection{Running the Code}
Once {\tt eos.parameters} and {\tt eos.quantities} are present in your directory, you can run the code with {option `3'}
  \begin{quotation}
    \noindent
    {\tt
      ./compose  \includegraphics[scale=1]{return.pdf}\\
      3  \includegraphics[scale=1]{return.pdf}\\
    }
  \end{quotation}
\vspace{-.4cm}
This should generate in particular a file {\tt eos.table}. The first
three columns contain the thermodynamic grid in the order $T$ (MeV),
$n_B$ (fm$^{-3}$), and $Y_q$ with the followings columns providing the quantities
specified in {\tt eos.quantities} in the same order and recalled by the terminal output. 

Up to seven other files are generated, {\tt eos.init}, {\tt eos.report},
  {\tt eos.beta}, {\tt eos.errdistr}, {\tt eos.info.json}, {\tt eos.nb.ns}, and {\tt eos.thermo.ns}.
These files are usually not of interest for the first-time user, see
the full manual for details.

\subsection{Examples}
Download the example files from the \textsc{CompOSE} web page `software' section. They contain four distinct examples for running the {\tt compose} code. 

\subsubsection{Generating a Table for $\beta$-Equilibrated Matter from a General Purpose EoS}

The following steps will allow you to run the {\tt compose} code and
generate a table containing the energy per baryon, pressure, and scaled
baryon chemical potential ($\mu_B - m_n$) for $\beta$-equilibrated
matter at a temperature of $T = 100$ keV. This table can be used for
solving the equations describing cold $\beta$-equilibrated neutron
stars. The example has been obtained with the FOP(SFHoY) EoS, {see} 
 \url{http://compose.obspm.fr/eos/118/}. Download {\tt eos-beta.zip} from the software section and unpack it. {Then enter,}
\begin{quotation}
\noindent {\tt cp eos.parameters.beta eos.parameters}  \includegraphics[scale=1]{return.pdf}\\
{\tt cp eos.quantities.beta eos.quantities}  \includegraphics[scale=1]{return.pdf}\\
{\tt ./compose} \includegraphics[scale=1]{return.pdf}\\
{\tt 3}  \includegraphics[scale=1]{return.pdf}\\
\end{quotation}
\vspace{-.4cm}
You can find a snapshot when running {\tt compose} for this particular case in Appendix \ref{app:beta}.

\begin{figure}[H]
    \includegraphics[width=0.52\textwidth]{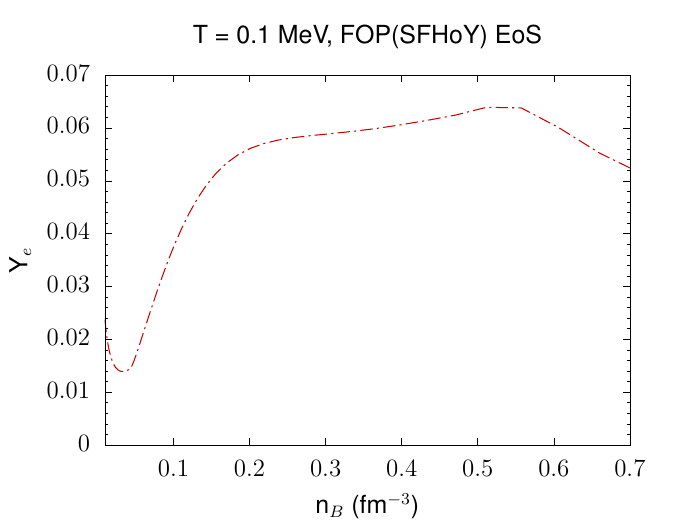}
  \caption{The electron fraction in $\beta$-equilibrium for the FOP(SFHoY) EoS at $T = 0.1$ MeV as function of baryon number density. \label{fig1}}
\end{figure}

The file {\tt eos.table} should then contain the {columns} 
\begin{quotation}
\mbox{}\hspace{-10mm}
\begin{tabular}{cccccc}
  $T$ & $n_{B}$ & $Y_{q}$ & $\mathcal{E}$ & $p$ & $\mu_{B}-m_{n}$ \\
  \mbox{} [MeV] & [fm$^{-3}$] & & [MeV] & [MeV~fm$^{-3}$] & [MeV] 
\end{tabular}
\end{quotation}
for $\beta$-equilibrated matter, where the units are given in parentheses. You can compare the output with the
provided data in {\tt eos.table.beta}. The result for the electron fraction as function of baryon number density is shown in Figure~\ref{fig1}. 

Please note that $\beta$-equilibrium is defined here via vanishing
lepton chemical potential and obtained by simple root finding. At
finite temperature or for trapped neutrinos, it does not necessarily
correspond to the physical $\beta$-equilibrium.

You can run the present example with any general purpose table
containing electrons. Please note that potentially you have to adapt
the temperature to the lowest entry in the respective table and the
range in baryon number density in the file {\tt eos.parameters}
in lines 8 (minimum values, order $T, n_B, Y_e$) and 9 (maximum
values, order $T,n_B, Y_e$). Output is generated in this case only if
a solution for $\beta$-equilibrium is found within the ranges in $Y_e$
of the table.

\subsubsection{Generating a Table at Fixed Entropy per Baryon from a General Purpose EoS}

The following steps will allow you to run the {\tt compose} code and
generate a table containing the energy par baryon, pressure, and scaled
baryon chemical potential ($\mu_B - m_n$) at a constant fixed entropy
per baryon, $s = 2 k_B$ and $Y_e = 0.4$, as function of baryon number density.  The example has been obtained with the OMHN(DD2Y) EoS,
{see} 
 \mbox{\url{http://compose.obspm.fr/eos/104/}}. Download {\tt eos-s.zip}
from the software section and unpack it. {Then enter,}
\begin{quotation}
\noindent {\tt cp eos.parameters.s eos.parameters} \includegraphics[scale=1]{return.pdf}\\
{\tt cp eos.quantities.s eos.quantities} \includegraphics[scale=1]{return.pdf}\\
{\tt ./compose} \includegraphics[scale=1]{return.pdf}\\
{\tt 3} \includegraphics[scale=1]{return.pdf}\\
\end{quotation}
\vspace{-.4cm}
You can find a snapshot when running {\tt compose} for this particular case in Appendix \ref{app:s}.

\begin{figure}[H]
    \includegraphics[width=0.52\textwidth]{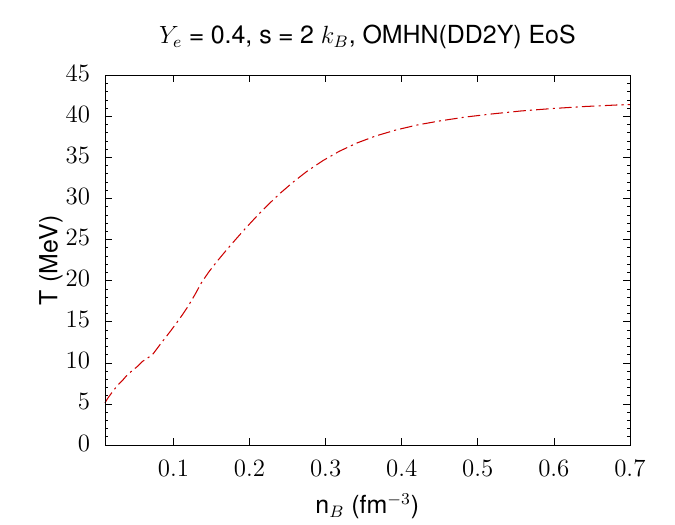}
  \caption{The temperature at constant entropy per baryon $s = 2 k_B$ and electron fraction $Y_e = 0.4$ for the OMHN(DD2Y) EoS as function of baryon number density. \label{fig2}}
\end{figure}

The file {\tt eos.table} should then contain the {columns}
\begin{quotation}
\mbox{}\hspace{-10mm}
\begin{tabular}{ccccccc}
  $T$ & $n_{B}$ & $Y_{q}$ & $s$ & $\mathcal{E}$ & $p$ & $\mu_{B}-m_{n}$ \\
  \mbox{} [MeV] & [fm$^{-3}$] & & [$k_{B}$] & [MeV] & [MeV~fm$^{-3}$] & [MeV]
\end{tabular}
\end{quotation}
with the units given in parentheses.
Figure~\ref{fig2} displays the result for the temperature as
function of baryon number density.  You can compare the ouput with the
provided data in {\tt eos.table.s}. You can run the present example
with any general purpose table. If the option of fixed entropy per
baryon is chosen, the first entries in lines 8--11 of {\tt
  eos.parameters} concern the minimum and maximum value of $s$, the
number of points, and logarithmic/linear scaling in entropy. Since the
EoS tables are generated as a function of temperature, $s_\mathit{min}$
and $s_\mathit{max}$ depend on baryon number density and hadronic
charge fraction. Output is generated in this case only if a solution
for the given value of fixed entropy is found within the ranges in
temperature of the table for the given value of $n_B$ and $Y_{q}$.

\subsubsection{Extracting Composition Information from a General Purpose Table}

The following steps will allow you to run the {\tt compose} code and
generate a table containing the energy per baryon and pressure, as
well as all particle fractions as a function of temperature for a
general purpose EoS. In the present example, particle fractions for
electrons, nucleons, and hyperons are listed, as well as light nuclei
and one average heavy nucleus. The Dirac effective mass for neutrons
is given too. The example has been obtained with the FOP(SFHoY) EoS, {see} 
\url{http://compose.obspm.fr/eos/118/}. If you did not yet download
it, get {\tt eos-3d.zip} from the software section and unpack it. {Then enter,}
\begin{quotation}
\noindent {\tt cp eos.parameters.3d eos.parameters} \includegraphics[scale=1]{return.pdf}\\
{\tt cp eos.quantities.3d eos.quantities} \includegraphics[scale=1]{return.pdf}\\
{\tt ./compose} \includegraphics[scale=1]{return.pdf}\\
{\tt 3} \includegraphics[scale=1]{return.pdf}\\
\end{quotation}
\vspace{-.4cm}
You can find a snapshot when running {\tt compose} for this particular case in Appendix \ref{app:3d}.

The file {\tt eos.table} should then contain in the columns
\begin{quotation}
\mbox{}\hspace{-10mm}
  \begin{tabular}{ccccccccccc}
  $T$ & $n_{B}$ & $Y_{q}$ & $\mathcal{E}$ & $p$ & $\{Y_{i}\}$
  & $Y_{av}$ & $A_{av}$ & $Z_{av}$ & $N_{av}$ & $\frac{m^{D}_{n}}{m_{n}}$  \\
  \mbox{} [MeV] & [fm$^{-3}$] & & [MeV] & [MeV~fm$^{-3}$] & & & & & & 
\end{tabular}
\end{quotation}
at $Y_e = 0.3$ and $n_B = 0.01$ fm$^{-3}$ for different values of
the temperature. The symbol $\{Y_{i}\}$ denotes the occurrence of 
$12$ columns of this quantity, in the order $i=n$, $p$, $\Lambda$, $\Sigma^-$, $\Sigma^0$, $\Sigma^+$, $\Xi^-$, $\Xi^0$, ${}^4\mathrm{He}$, ${}^3\mathrm{He}$, ${}^3\mathrm{H}$, ${}^2\mathrm{H}$.

\begin{figure}[H]
    \includegraphics[width=0.52\textwidth]{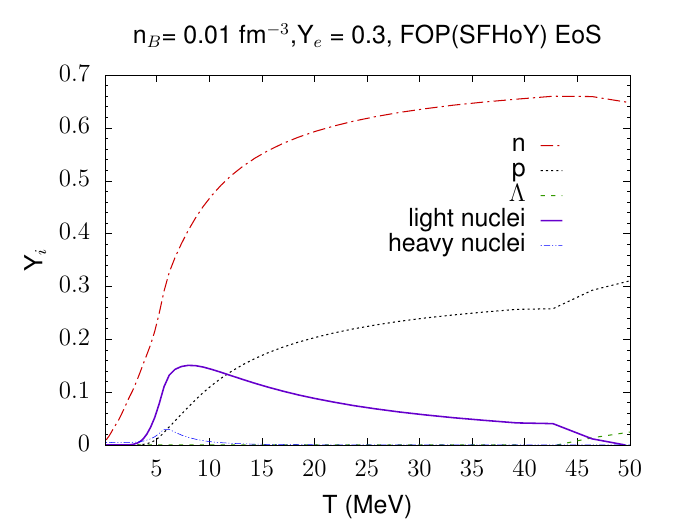}
  \caption{Different particle fractions as function of temperature for the FOP(SFHoY) EoS at $n_B$~=~0.01~fm$^{-3}$ and $Y_e = 0.3$. \label{fig3}}
\end{figure}

You can compare the output with the provided data in {\tt
  eos.table.3d}, from which some particle fractions are shown in
Figure~\ref{fig3}. You can run the present example with any general
purpose table. Please note that you will potentially need to adapt the
temperature to the lowest entry in the respective table and the range
in baryon number density in the file {\tt eos.parameters} in
lines 8 (minimum values) and 9 (maximum values). If you employ a table
not containing hyperons, then the number of pairs in line 6 has to be
adapted and the indices starting with 100 in line 8 have to be
removed. In the same way, if your table does not contain information
about individual nuclei, then in line 6 the number of pairs has to be
adapted and the four-digit entries in line 8 have to be removed. If
the Dirac effective mass is not available, line 10 should contain a 0
and line 12 should be empty. If you run {\tt compose} with task `1',
you will be guided through the generation of a new {\tt
  eos.quantities} file.
  
  \subsubsection{Extracting Sound Speed and Adiabatic Index from a Cold Neutron Star Table}

The following steps will allow you to run the {\tt compose} code and
generate a table containing the energy per baryon, pressure, and scaled
baryon chemical potential ($\mu_B - m_n$), as well as squared speed of sound, adiabatic index, and entropy per baryon as a function of baryon number density for a cold neutron star EoS. The example has been obtained with the RG(SkA) EoS, {see} 
 \url{http://compose.obspm.fr/eos/96/}. Download  {\tt eos-ns.zip} from the software section and unpack it. {Then enter,}
 \clearpage
\begin{quotation}
\noindent {\tt cp eos.parameters.ns eos.parameters} \includegraphics[scale=1]{return.pdf}\\
{\tt cp eos.quantities.ns eos.quantities} \includegraphics[scale=1]{return.pdf}\\
{\tt ./compose} \includegraphics[scale=1]{return.pdf}\\
{\tt 3} \includegraphics[scale=1]{return.pdf}\\
\end{quotation}
\vspace{-.4cm}
 You can find a snapshot when running {\tt compose} for this particular case in Appendix \ref{app:ns}.

\begin{figure}[H]
    \includegraphics[width=0.52\textwidth]{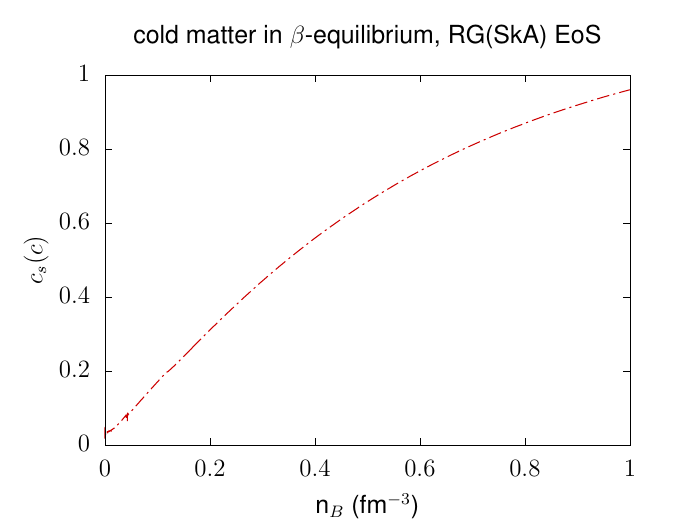}
  \caption{Sound speed as function of baryon number density for the cold $\beta$-equilibrated RG(SkA)~EoS. \label{fig4}}
\end{figure}

The file {\tt eos.table} should then contain the columns
\begin{quotation}
\mbox{}\hspace{-10mm}
\begin{tabular}{ccccccccc}
  $T$ & $n_{B}$ & $Y_{q}$ & $\mathcal{E}$ & $p$ & $\mu_{B}-m_{n}$ & $c_{s}^{2}$ & $\Gamma$ & s\\
  \mbox{} [MeV] & [fm$^{-3}$] & & [MeV] & [MeV~fm$^{-3}$] & [MeV] & [$c^{2}$] & & [$k_{B}$]
\end{tabular}
\end{quotation}
at $T = 0$ MeV for $\beta$-equilibrated matter. You can compare
the output with the provided data in {\tt eos.table.ns}, the result
for the sound speed is shown in Figure~\ref{fig4}. You can run the
present example with any cold neutron star table. Please note that
you will potentially need to adapt $n_B$ to the lowest entry in the
respective table in the file {\tt eos.parameters} in line 8 and the
maximum value in line 9.

\vspace{-6pt} 
\authorcontributions{Set up of the data base and compose software---M.O. and S.T.; web site and web tools---M.M. and M.O.; writing---V.D. and L.T.; review and editing---V.D., M.M., M.O., C.P., L.T., and S.T. All authors have read and agreed to the published version of the manuscript.}

\funding{\textsc{CompOSE} would not be possible without the financial and organisatorial
support from a large number of institutions and individual contributors.
We gratefully acknowledge support by the CompStar network of the European Science Foundation (ESF), the birthplace of
the \textsc{CompOSE} project, the COST actions NewCompStar, and PHAROS, 
a grant from the Polish Ministry for
Science and Higher Education (MNiSW) supporting the ``CompStar'' activity,
by the Instytut Fizyki Teoretycznej of the Uniwersytet Wroc\l{}awski,
the National Science Centre Poland
(Narodowe Centrum Nauki, NCN) within the ``Maestro'' programme under
contract No. DEC-2011/02/A/ST2/00306, by the
``hadronphysics3'' network within the seventh framework program of the
European Union,
by the GSI Helmholtzzentrum f\"{u}r Schwerionenforschung GmbH,
by the Helmholtz International
Center for FAIR within the framework of the LOEWE program launched
by the state of Hesse via the Technical University Darmstadt,
by the Helmholtz Association (HGF) through the Nuclear Astrophysics
Virtual Institute (VH-VI-417),
by the ExtreMe Matter Institute EMMI in the framework
of the Helmholtz Alliance "Cosmic Matter in the Laboratory",
by the DFG cluster of excellence ``Origin and Structure of
the Universe'',
by the DFG through grant SFB~1245,
by the SN2NS project ANR-10-BLAN-0503. 
{L.T. also acknowledges} 
 support from CEX2020-001058-M (Unidad de Excelencia ``Mar\'{\i}a de Maeztu"), PID2019-110165GB-I00 financed by the spanish MCIN/ AEI/10.13039/501100011033/, as well as by the EU STRONG-2020 project, under the program  H2020-INFRAIA-2018-1 grant agreement no. 824093, and the CRC-TR 211 "Strong-interaction matter under extreme conditions"- project No. 315477589---TRR 211. 
V.~D. acknowledges support from the National Science Foundation
under grants PHY1748621, MUSES OAC-2103680, and NP3M
PHY-2116686. C.~P. acknowledges support from the Funda\c{c}\~ao para a Ci\^encia e Tecnologia  under the Projects UIDP/04564/2020 and UIDB/04564/2020.}

\dataavailability{All data are publicly {available on} 
 \url{https://compose.obspm.fr}.} 

\acknowledgments{We thank Jean-Yves Giot for creating the first version of the \textsc{CompOSE} web site.}

\conflictsofinterest{{The authors declare no conflict of interest.}}

\appendixtitles{yes} 
\appendixstart
\appendix
\section[\appendixname~\thesection. Notations]{Notations}
\label{app:a}

\begin{table}[H]
\caption{{Notations used in the Quick Guides} 
}
  \newcolumntype{C}{>{\centering\arraybackslash}X}
\begin{tabularx}{\textwidth}{Cm{5cm}<{\centering}C}
\toprule 
    \textbf{Symbol}& \textbf{Quantity} & \textbf{Unit} \\
    \midrule
    $T$ & temperature & MeV \\
    $n_{B}$ & baryon number density & fm${}^{-3}$ \\
    $Y_{i}$ & number density fraction of particle $i$ & -- \\
    $Y_{q}$ & hadronic (and quark) charge fraction & -- \\
    $p$ & pressure & MeV~fm${}^{-3}$ \\
    $\mathcal{E}$ & energy per baryon & MeV \\
    $m_{n}$ & neutron mass & MeV \\
    $\mu_{B}$ & baryon chemical potential & MeV \\
    $c_{s}$ & speed of sound & $c$ \\
    $s$ & entropy per baryon & $k_{B}$\\
    \bottomrule
\end{tabularx}
\end{table}

\section[\appendixname~\thesection. Example of a Table for \boldmath{$\beta$}-Equilibrated Matter from a General Purpose EoS]{Example of a Table for \boldmath{$\beta$}-Equilibrated Matter from a General Purpose EoS}
\label{app:beta}

\begin{figure}[H]
  \fbox{
    \includegraphics[width=0.55\textheight]{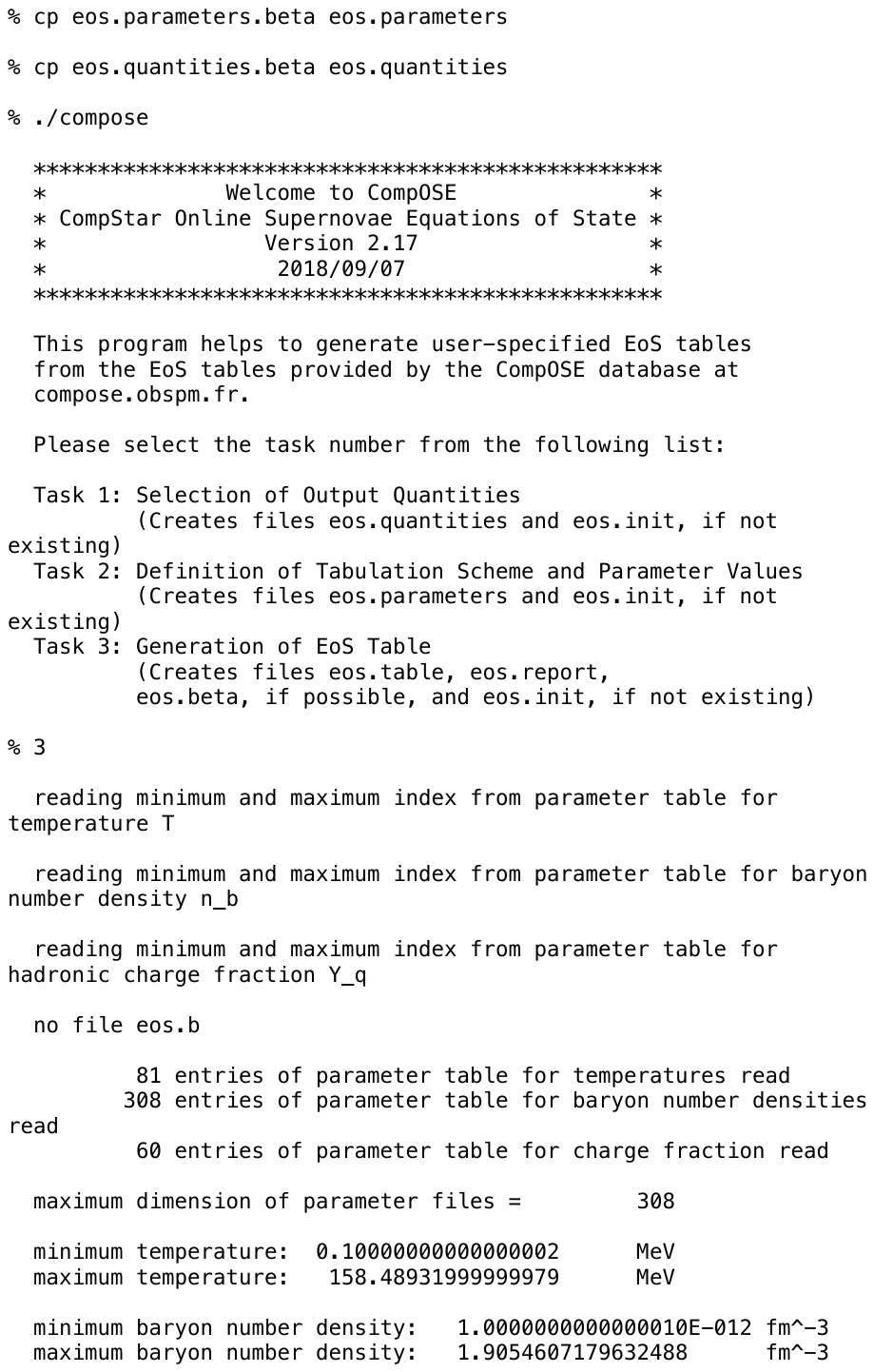}
    }
  \caption{{Snapshot when running compose for generating a table for $\beta$-equilibrated matter from a general purpose EoS.}
}
\end{figure}

\section[\appendixname~\thesection. Example of  a Table at Fixed Entropy per Baryon from a General Purpose~EoS]{Example of  a Table at Fixed Entropy per Baryon from a General Purpose~EoS}
\label{app:s}

\begin{figure}[H]
  \fbox{
    \includegraphics[width=0.55\textheight]{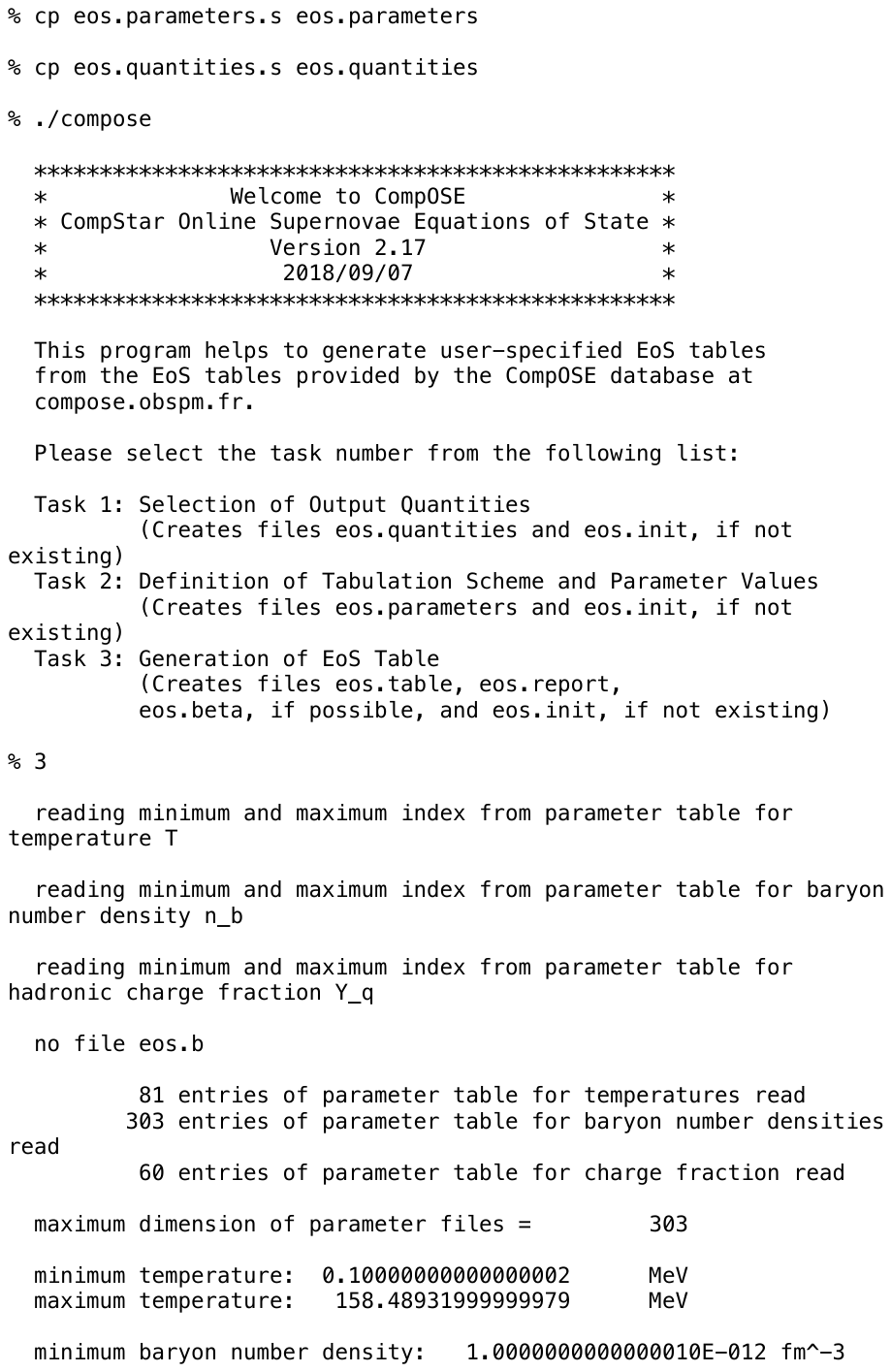}
    }
    \caption{{Snapshot when running compose for generating a table at fixed entropy per baryon from a general purpose EoS }} 
\end{figure}

\section[\appendixname~\thesection. Example  of the Composition Information from a General Purpose Table]{Example  of the Composition Information from a General Purpose Table}
\label{app:3d}

\begin{figure}[H]
  \fbox{
    \includegraphics[width=0.55\textheight]{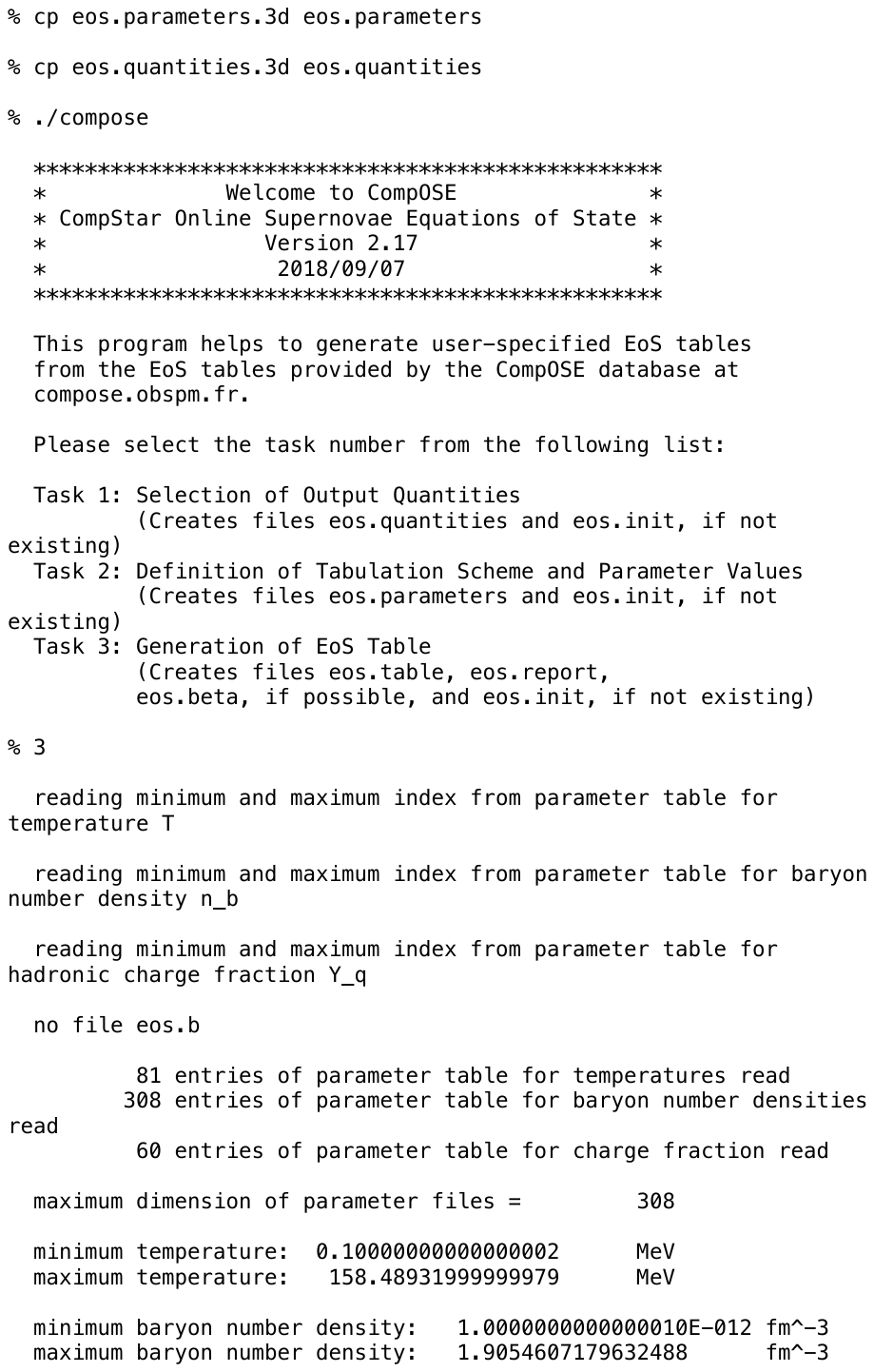}
    }
     \caption{{Snapshot when running compose for extracting compositional information from a general purpose table}} 
\end{figure}

\section[\appendixname~\thesection. Example of the Sound Speed and Adiabatic Index from a Cold Neutron Star Table]{Example of the Sound Speed and Adiabatic Index from a Cold Neutron Star Table}
\label{app:ns}

\begin{figure}[H]
  \fbox{
    \includegraphics[width=0.55\textheight]{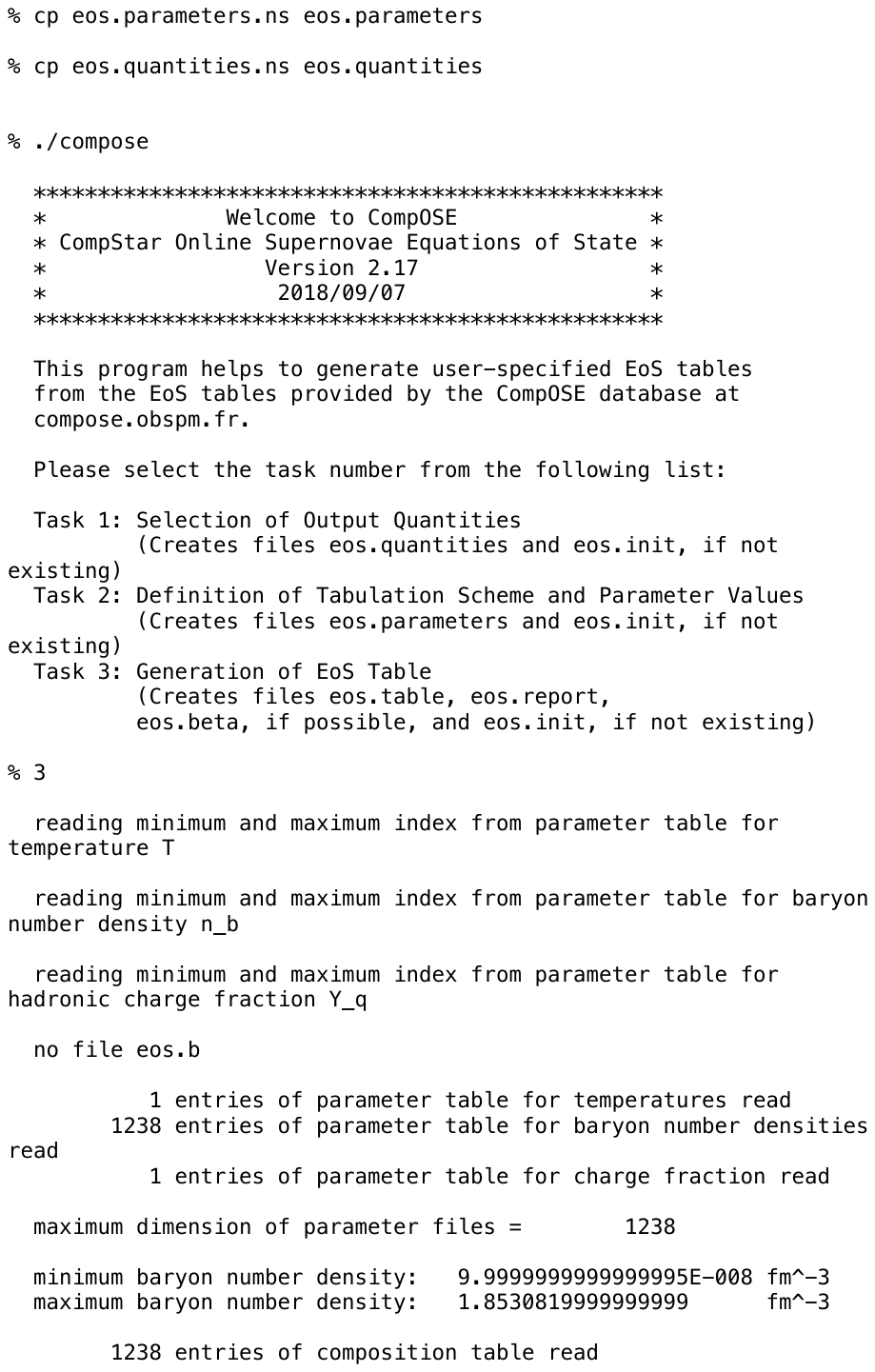}
    }
    \caption{{Snapshot when running compose for computing sound speed and adiabatic index from a cold neutron star EoS}} 
\end{figure}


\begin{adjustwidth}{-\extralength}{0cm}
\reftitle{References}




\end{adjustwidth}
\end{document}